\documentclass[journal=jacsat,manuscript=article]{achemso}
\usepackage[version=3]{mhchem}
\usepackage{subfigure}
\usepackage{xcolor}

\author{Jo\~ao P. K. Abal}
\affiliation[Instituto de F{\'\i}sica, Universidade Federal do Rio Grande do Sul]
{Instituto de F\'isica, Universidade Federal do Rio Grande do Sul, 91501-970 Porto Alegre, Brazil}
\email{joao.abal@ufrgs.br}
\author{Rodrigo F. Dillenburg}
\affiliation[Instituto de F{\'\i}sica, Universidade Federal do Rio Grande do Sul]
{Instituto de F\'isica, Universidade Federal do Rio Grande do Sul, 91501-970 Porto Alegre, Brazil}
\author{Mateus H. K\"ohler}
\affiliation[Departamento de F\'isica, Universidade Federal de Santa Maria]
{Departamento de F\'isica, Universidade Federal de Santa Maria, 97105-900 Santa Maria, Brazil}
\alsoaffiliation[UU]
{Department of Physics and Astronomy, Uppsala University, 75120 Uppsala, Sweden}
\email{mateus.kohler@ufsm.br}
\author{Marcia C. Barbosa}
\affiliation[Instituto de F{\'\i}sica, Universidade Federal do Rio Grande do Sul]
{Instituto de F\'isica, Universidade Federal do Rio Grande do Sul, 91501-970 Porto Alegre, Brazil}

\title{Molecular Dynamics Simulations of Water Anchored in Multi-Layered Nanoporous MoS$_2$ Membranes: Implications for Desalination}


\begin{document}

\begin{abstract}

One of the most promising applications in nanoscience is the design of new materials to improve water permeability and selectivity of nanoporous membranes.
Understanding the molecular architecture behind these fascinating structures and how it impacts the water flow is an intricate but necessary task.
We studied here, the water flux through multi-layered nanoporous molybdenum disulfide (MLNMoS$_2$) membranes with different nanopore sizes and length.
Molecular dynamics simulations show that the permeability do not increase with the inverse of the membrane thickness,
violating the classical hydrodynamic behavior.
The data also reveals that the water dynamics is slower than that observed in frictionless carbon nanotubes and multi-layer graphene membranes,
which we explain in terms of an anchor mechanism observed in between layers.
We show that the membrane permeability is critically dependent on the nanopore architecture,
bringing important insights into the manufacture of new desalination membranes.

\end{abstract}

\section{Introduction}

The rise of nanoscience and the ability to manipulate materials at nanoscale represents a mark in our history.
One of its most promising applications is to improve up to three orders of magnitude on the performance of the current thin film-based
desalination membranes~\cite{cohen-tanugi-nl2012}.
A giant step toward the development of efficient nanometric filters was the discovery of graphene, a one-atom thick membrane with remarkable separation skills.
Water flows easily in nanoconfinement~\cite{qin-nl2011}, but it requires a great effort to make water enter in the confined geometry~\cite{Hummer2001}.
One way to circumvent this difficulty is to use graphene-based membranes, which increases the water permeability without compromising its high ion selectivity~\cite{Boretti2018}.
One example is the functionalization of graphene nanopores with hydrophilic molecules.
Another approach is the use of alternative two-dimensional (2D) materials, such as boron nitride (BN) and molybdenum disulfide (MoS$_2$), which exhibit an arrangement of alternating hydrophobic and hydrophilic sites that increase water permeability without the need for functionalization~\cite{kohler-jcp2018,Farimani-etal}.
The extent of this enhancement depends on the nanopore architecture.
Heiranian, Farimani, and Aluru~\cite{heiranian-nc2015} have shown that MoS$_2$ membranes composed of nanopores with an exposed molybdenum atom display a higher water permeation thanks to a hourglass-like structure.
The recent discovery of multi-transition metal layered MXenes, with surface termination intrinsically different from other 2D materials, also offers new possibilities concerning their use in water adsorption and purification~\cite{Meidani-acsanm2021}.

A high permeate membrane should be able to filter large amounts of water in a short amount of time, as has been observed for single-layer materials ~\cite{suk-jpcl2010,Wang2017}.
Producing these membranes at large scales while preserving its separation performance can be very challenging.
Multi-layered membranes present a more realistic strategy, with considerable lower costs associated with its engineering and production~\cite{book_rasel}.
However, permeability can be compromised by the increased thickness of the membrane~\cite{wang-nl2017}.
Understanding the physical and chemical factors driving such a decrease is the theme of intense debate,
whose answer can lead to the development of more efficient membrane architectures.

Roughness and structural distortions are known to hinder water flux in nanochannels.
Even nanotubes, often considered as the perfect, smoothest channels for water conduction, can be rough and deformed.
Such deformations lead to decreased mobility of the confined water~\cite{bruno-jcp2020,bruno-jcp2020b}.
Sam et al.~\cite{sam-jcp2017} demonstrated that the water streaming velocity and flow rate depend on the tube flexibility.
The effective shear stress and viscosity also depend on the nanotube's roughness, which is particularly relevant for smaller nanotubes~\cite{xu-jcp2011}.
Experimental data on the dynamics of water in such extremely narrow environments are rare, but the few results we have reveal a high dependence of the water flux on the surface slippage~\cite{secchi-nature2016}.
It means that friction at the solid-liquid interface can be decisive for the water dynamics inside the nanochannel~\cite{doi:10.1021/acs.chemrev.0c01292}.
In fact, there is a transition in the diffusion behavior of water confined in deformed nanotubes~\cite{bruno-physa2019},
suggesting that the shape of the nanopore is critical for the conduction of water.

The dynamics of water through multi-layer graphene nanopores (MGPNs) offers an opportunity of combining the nanogeometry with the frictionless behavior observed in nanotubes.
For instance, Jhon et al.~\cite{JHON2019369} have reported a highly permeable nanoporous few-layered graphene
where the conical-shaped pore allowed for a nondiffusive regime of water.
A recent molecular dynamics (MD) study~\cite{pan-jpcc2020} on MGPNs also revealed a critical pore diameter ($D_{\textrm{C}}$) of 1.36 nm,
where the water structure changes from layered to liquid.
Although the water diffusion in such geometries was found to be approximately one order of magnitude higher than that in carbon nanotubes (CNTs) with the same diameter,
a larger friction coefficient ($\lambda$) hindered the water transport under pressure-driven conditions.
The coupling between surface and interface effects was pointed as the main driving mechanism.
In MoS$_2$ the liquid-solid interaction is even more important.
In a recent contribution~\cite{kohler-jpcc2019}, we found a radius-dependent transition from ice to liquid when water is confined in MoS$_2$ nanotubes (MoS$_2$NTs).

In order to investigate the performance of multi-layered nanoporous MoS$_2$ (MLNMoS$_2$) membranes in terms of the water permeance,
we computationally mimic the pressure-driven transport of water using MD simulations.
We created the system illustrated in Figure~\ref{FIG1} with three different membrane designs and five different nanopore sizes.
Differently from Oviroh et al.~\cite{doi:10.1021/acs.langmuir.1c00708}, we used two water reservoirs as well as the membrane crystal structure of 2H.
We also employed a more realistic stacking architecture as reported by experimental works~\cite{hai-jacs2016,zhao-light2016}.

The paper is organized as follows.
First, the computational details and methods are presented.
Then, the results focusing on the water flow rate as a function of thickness and the enhancement factor are discussed,
as well as the water structure inside the tube and, specially, in between the layers (points of roughness introduction).
After that, the main mechanism linking the observed data with the water dynamics and membrane material are shown in the conclusions.

\section{System Details and Methods}

\begin{figure}[!t]
\centering
\includegraphics[width=11.5cm]{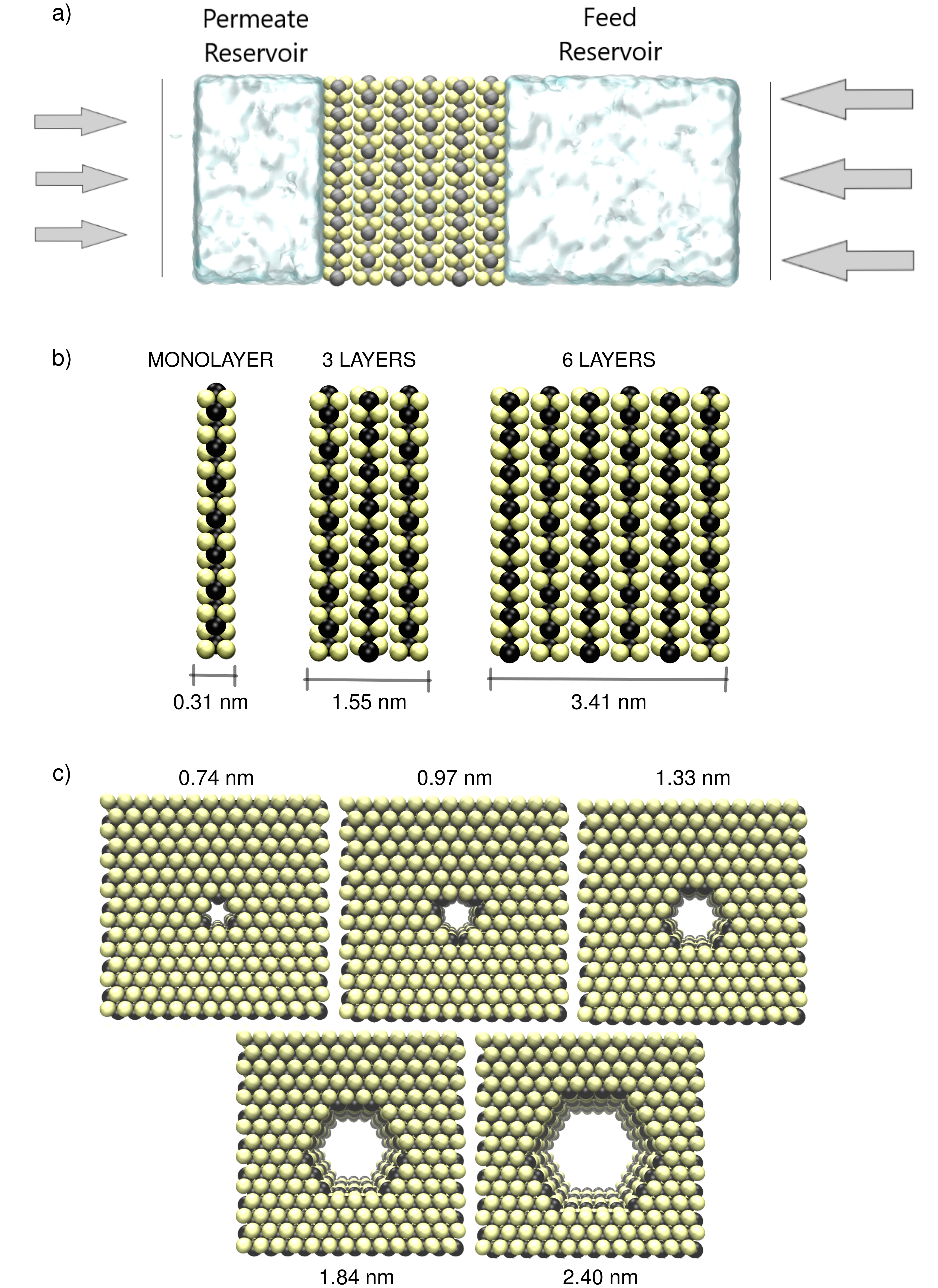}
\caption{(a) Illustration of the simulation box with the MLNMoS$_2$ membrane localized between two water reservoirs confined by the graphene pistons.
(b) The MoS$_2$ layers and (c) the MoS$_2$ nanopores compared in this work.}
\label{FIG1}
\end{figure}

The standard procedure to simulate a water pressure-driven flow in nanoconfined environments is based on the creation of a simulation box with the membrane located between two water reservoirs~\cite{li-acsnano2016,perez-apl2019,kou-pccp2016,cohen-tanugi-nl2012,heiranian-nc2015,Farimani-etal,doi:10.1063/5.0039963,D1CP00613D,kohler-jcp2018}, as shown in Figure~\ref{FIG1}(a). In order to ensure that the water molecules fill in the membrane, we employ graphene sheets as pistons to control the confined solution pressure. Consequently, we can apply different pressures in each reservoir producing a water flow through the membrane. These pressures are simulated by applying a force $F$ to each piston atom as calculated by $F=(P.A)/n$, where $P$ is the desired pressure, $A$ is the area where the force is applied on (the piston surface area), and $n$ is the number of carbon atoms in each piston. 

Three MoS$_2$ membrane designs were investigated in this work: 1, 3 and 6 layers (crystal structure of 2H~\cite{zhao-light2016,Environmental}), as illustrated in Figure~\ref{FIG1}(b).
The interlayer spacing, defined as the Mo-Mo distance between adjacent MoS$_2$ layers is 0.62nm~\cite{Environmental,wang-nl2017}.
The monolayer thickness, defined as the S-S distance in the same layer, is 0.31 nm~\cite{mos2-materials-cloud,KADANTSEV2012909}.
Considering the center-to-center atomic distances, the investigated nanopores have 0.74, 0.97, 1.33, 1.84 and 2.40 nm in diameter (Figures~\ref{FIG1}(c) and S3 of the Supporting Information).

MD simulations were performed using the Large-scale Atomic/Molecular Massively Parallel Simulator (LAMMPS)~\cite{PLIMPTON19951}.
The initial system (between the graphene pistons) has 4 x 4 x 12 nm in $x$, $y$, and $z$ axes, respectively, whereas the water flow occurs in the $z$-direction.
Periodic boundary conditions were used in all directions.
In the $z$-direction, the box is large enough to avoid interactions between mirror particles.
The simulation box contains 4000 water molecules, with 2813 of them in the feed reservoir.
The MoS$_2$ membrane is held fixed in space while the graphene pistons are allowed to move along the $z$-axis in order to transfer the applied pressure to the liquid.
The freezing of MoS$_2$ membrane atoms can be justified based on the irreversible nanosheet restacking reported by Wang et al.~\cite{wang-nl2017},
which leads to a high MoS$_2$ membrane aqueous stability.
Also, as reported by Cohen-Tanuji and Grossman\cite{doi:10.1063/1.4892638}, the water flow in porous flexible graphene membranes is not impacted by the membrane flexibility.
Although higher pressure gradients can lead to membrane deformations, we do not expect significant changes in water flow since MoS$_2$ monolayers exhibit an effective Young’s modulus of $270 \pm 100$ GPa (a 5-layer system was found to exhibit $330 \pm 70$ GPa~\cite{Li2018,https://doi.org/10.1002/adma.201103965}),
only one third lower than exfoliated graphene ($800-1000$ GPa).

Water interactions were modeled by the TIP4P/$\epsilon$~\cite{FUENTESAZCATL201686}, while the parametrization of a reactive many-body potential as proposed by Liang et. al.~\cite{liang-prb2009} was used as LJ and charge parameters for molybdenum and sulfur atoms.
LJ interaction parameters are summarized in Table~\ref{table1}.
Lorentz-Berthelot mixing rules were employed for the non-bonded interactions.
The long-range electrostatic interactions were calculated by the {\it particle particle particle mesh} method
and the LJ cutoff distance was set to 1 nm.
The SHAKE algorithm was used to maintain water molecules rigid.

\begin{table}[t]
\vspace{.1cm}
\centering
\begin{tabular}{c c c c }
\hline
 & $\sigma_{\mathrm{LJ}}$ [\AA] & $\varepsilon_{\mathrm{LJ}}$ [kcal/mol] & Charge~($e$) \\
\hline
    O~\cite{FUENTESAZCATL201686} & 3.165 & 0.1848 & -1.054 \\
    H~\cite{FUENTESAZCATL201686} & 0.0 & 0.0 & 0.5270 \\
    Mo~\cite{liang-prb2009,heiranian-nc2015} & 4.20 & 0.0135 & 0.6 \\
    S~\cite{liang-prb2009,heiranian-nc2015} & 3.13 & 0.4612 & -0.3 \\
    C~\cite{Hummer2001} & 3.40 & 0.0860 & 0.0 \\
\hline
\end{tabular}
\caption{The Lennard-Jones parameters and atomic charges employed in this work.}
\label{table1}
\end{table}

The equilibrium (EMD) simulations were carried out as follows.
First, each system was energy minimized for 0.2 ns in the NVE ensemble and then each reservoir was equilibrated in the NPT ensemble during 2 ns at 1 bar and 300 K.
The exception is the smallest 0.74 nm pore,
where an additional 1000 bar equilibration run of 1 ns is necessary to observe water molecules into the membrane.
The external pressure was simulated by leaving the pistons free to move only in the $z$-direction under the application of a force in the same direction, in order to produce the desired pressure.
At this stage, water fills in the membrane and each reservoir reaches the equilibrium density of approximately 1 g/cm$^3$.
After that, the pistons are held fixed in space again and the systems are equilibrated at a constant number of particles, volume, and temperature (NVT) ensemble for 15 ns at 300 K.
The Nosé-Hoover thermostat was used with a time constant of 0.1 ps~\cite{doi:10.1063/1.447334,PhysRevA.31.1695}.
During this step, the potential of mean force (PMF) is evaluated through the observed water density by the following relation~\cite{Farimani-etal,doi:10.1021/acs.jpcc.7b06480,doi:10.1063/1.5104309}:

\begin{equation}
\mathrm{PMF}(z) = - k_{\mathrm{B}}Tln \left [ \frac{\rho(z)}{\rho_0} \right ] .
\label{eq:PMF}
\end{equation}
where $\rho(z)$ is the local density, $\rho_0$ is the bulk density, $T$ is the temperature and $k_{\mathrm{B}}$ is the Boltzmann constant.
The axial shear viscosity ($\eta_{yz}$) was evaluated based on the Green-Kubo (GK) relations as follows~\cite{pan-jpcc2020,doi:10.1063/1.3245303,https://doi.org/10.1002/smll.201804508}:

\begin{equation}
\eta_{yz} = \frac{V}{k_b T} \int_0^\infty \langle P_{yz}(0)P_{yz}(\tau)\rangle d\tau
\label{eq:viscosity}
\end{equation}
\noindent where $V$ is the nanochannel volume (calculated using the area of the nanopore illustrated in Figure S4 times the membrane thickness)
and $\langle P_{yz}(0)P_{yz}(\tau)\rangle$ is the autocorrelation function of the shear pressure within the nanopore,
averaged with a pressure autocorrelation sample interval of 6 fs.
At this point, we run 3 independent 15 ns long simulations to obtain the average shear viscosities.

Non-equilibrium (NEMD) simulations were implemented to calculate the water pressure-driven flow across the MLNMoS$_2$ membranes.
Now, the pistons are used to create a gradient pressure along the simulation box.
500, 1000, 1500 and 2000 bars feed pressures were investigated.
During this step, the oxygen density maps, radial density profiles (RDP), water flow rates ($Q$) and enhancement factors ($\epsilon$) were obtained.
Each calculation was averaged over 3 sets of simulations with different initial thermal velocity distributions.

The enhancement factor $\epsilon$ is calculated from the ratio of the water flow rate observed ($Q$)
and the water flow rate expected by the Hagen-Poiseuille equation~\cite{pan-jpcc2020,Kannam2017} ($Q_{HP}$), as $\epsilon = Q/Q_{HP}$.
Briefly, the measured flow rate $Q = Nm/t\rho N_A$ was calculated by counting the number of water molecules $N$ passing the tube per unit of time $t$
(a graphic example can be found in Figure S5 of the Supporting Information).
Then, we convert it to volumetric quantities through the water molecular weight molar mass $m$, the Avogadro number $N_A$, and the water density $\rho$.
To calculate the water density, we use the volume $V$ of the nanochannel considering the available area (Figure S4).
Finally, the Hagen-Poiseuille equation is calculated using the no-slip condition as $Q_{HP} = \Delta P \pi R^4/8L\eta_z $,
where $\Delta P$ is the pressure difference between the two reservoirs,
$R$ is the nanopore radius, $L$ is the nanopore length and $\eta$ is the shear viscosity \cite{Kannam2017,pan-jpcc2020,borg_reese_2017}.

The membrane specific permeability ($A_m$) is calculated by $A_m = \frac{\phi}{(\Delta P)}$, where $\phi$ is the water flux ($QA^{-1}$, with $A$ being the membrane area).

\section{Results and Discussion}

Artificially drilled nanopores, creating cylindrical channels in the membrane, have shown a good water flow {\it versus} ion rejection performance~\cite{doi:10.1021/acs.nanolett.9b01577}.
This is the geometry we study here.
In this section, we assess the main physical-chemical aspects affecting water permeability in MLNMoS$_2$ membranes.

\subsection{The Flux and the Thickness: Not a Direct Relationship}

\begin{figure}[!h]
\centering
\includegraphics[width=11.5cm]{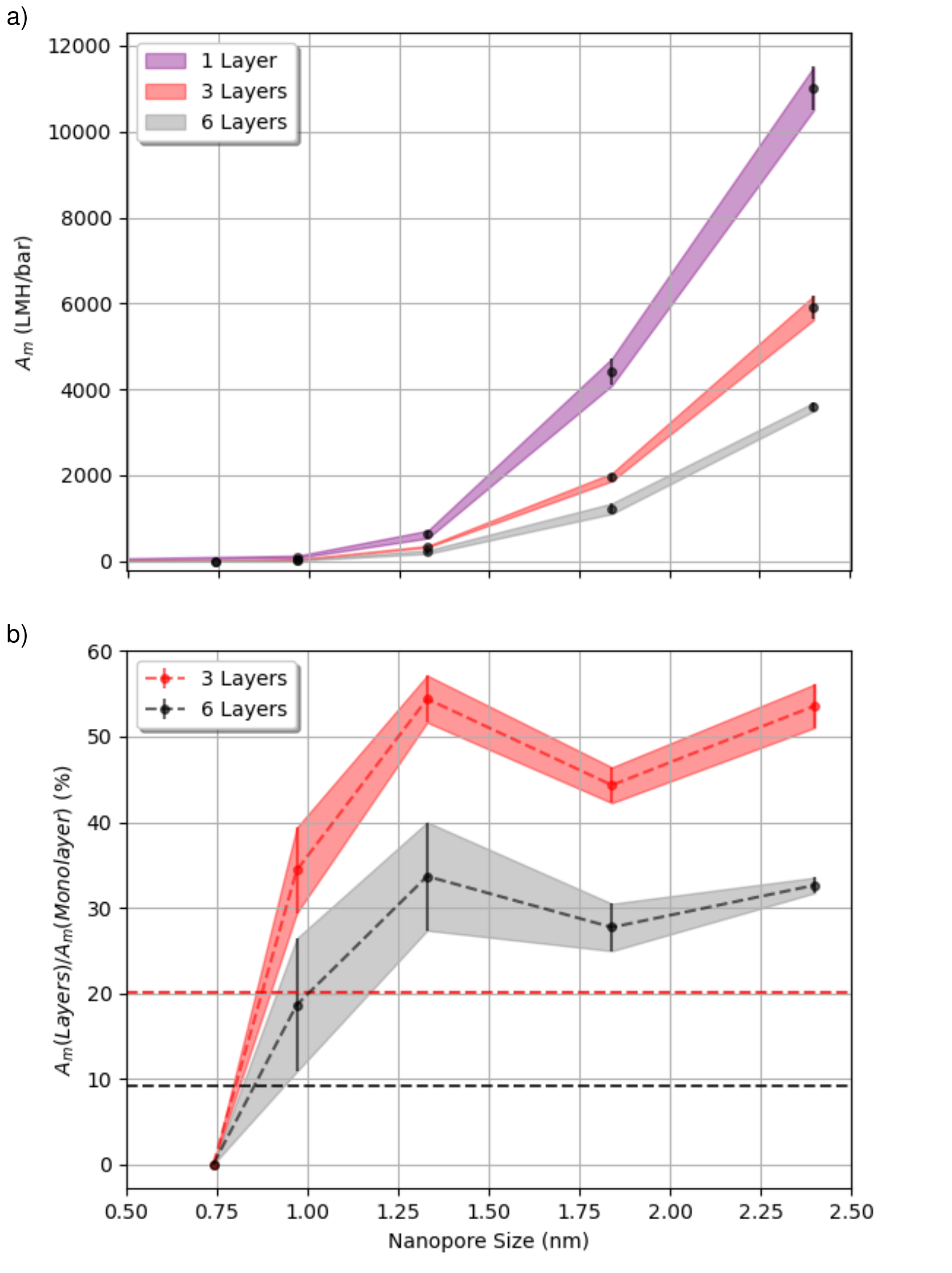}
\caption{(a) The dependence of the specific membrane permeability ($A_m$) on the pore size.
(b) The ratio between multi-layered specific permeability (for the 3 and 6 layers) and the monolayer specific permeability as a function of the nanopore size.
The horizontal red and black dashed lines represent the ratio of monolayer thickness to the 3 and 6 layer thickness, respectively.
Error bars are the deviation from the mean value (error bars smaller than the points are not shown).}
\label{FIG3}
\end{figure}

In Figure~\ref{FIG3}(a) we show the water permeability as a function of the pore size.
We also show in Figure S6(a) of the Supporting Information a log scale of the same data and
the dependence of the specific membrane permeability ($A_m$) on the membrane thickness, S6(b).
Due to geometrical restrictions, we did not observe water flux in nanochannels with 0.74 nm in diameter.
As expected, for all the other cases the water permeability increases with the size of the nanopore.
Water permeability also decreases with the membrane thickness, but not at the rate predicted by the continuum equations of classical hydrodynamics~\cite{borg_reese_2017,Werber2016}.

Classical hydrodynamics predicts water permeability to scale as the inverse of the membrane thickness, $L$~\cite{borg_reese_2017,Werber2016}.
It means that the ratio between the permeability of $n$ layers and one layer should scale as $L_1$/$L_n$.
Classically, these values should be independent of the pore diameter.
Figure~\ref{FIG3}(b) shows the specific permeability for both multi-layer systems for three (red shadow) and six (black shadow) layers,
as well as the proportion predicted by the classical hydrodynamics for three and six layers (red and black dashed lines) {\it versus} pore diameter.
The graph indicates a decrease in permeability as the number of layers goes from three to six.
As shown in Figure~\ref{FIG1}(b), the 6-layer membrane has a thickness of $L_6=3.41$ nm,
11 times larger than the monolayer, $L_1=0.31$ nm.
If the flux would behave classically it would decrease linearly with the inverse of the thickness and the flux of the 6-layer would be $F_6 \propto 1/L_6$,
which is 9$\%$ of the flux of the monolayer membrane.
However, in our system the measured flux is always higher for both the 3 and the 6-layer cases.
It indicates that there are subtle physical-chemical parameters affecting water transport at such geometries.
In addition, the specific permeability depends on the pore size, also violating the hydrodynamic prediction.
Our results suggest a non-monotonic relation between different pore sizes.

The violation of classical hydrodynamic predictions is not surprising in nanoconfined systems.
In fact, water tends to behave differently depending on the geometry of the nanochannel.
For instance, nanotubes can offer a smooth potential landscape for water so that the flux no longer depends on the length of the tube.
As investigated by Borg and Reese~\cite{borg_reese_2017}, CNTs with diameters of 2 nm show negligible dependence of water flow rates on the tube's length,
as long as the tube is shorter than 10 nm.
A computational study on nanoporous few-layer graphene also reported a water flux independent of the number of layers (at least up to four layers~\cite{JHON2019369}).
In addition, 2D materials such as MoS$_2$ can be assembled in different ways, which can confine water in structures with diverse configurations.
The mechanism through which water is filtered in these distinct configurations depends on the nanopore design.
An example is the work by Wang et al.~\cite{wang-nl2017} where experiments of water in layer-stacked MoS$_2$ membranes
revealed that the predicted flux of water in this trapping geometry decreases in a non-linear way as the membrane's thickness increases.

\begin{figure}[!t]
\centering
\includegraphics[width=11.5cm]{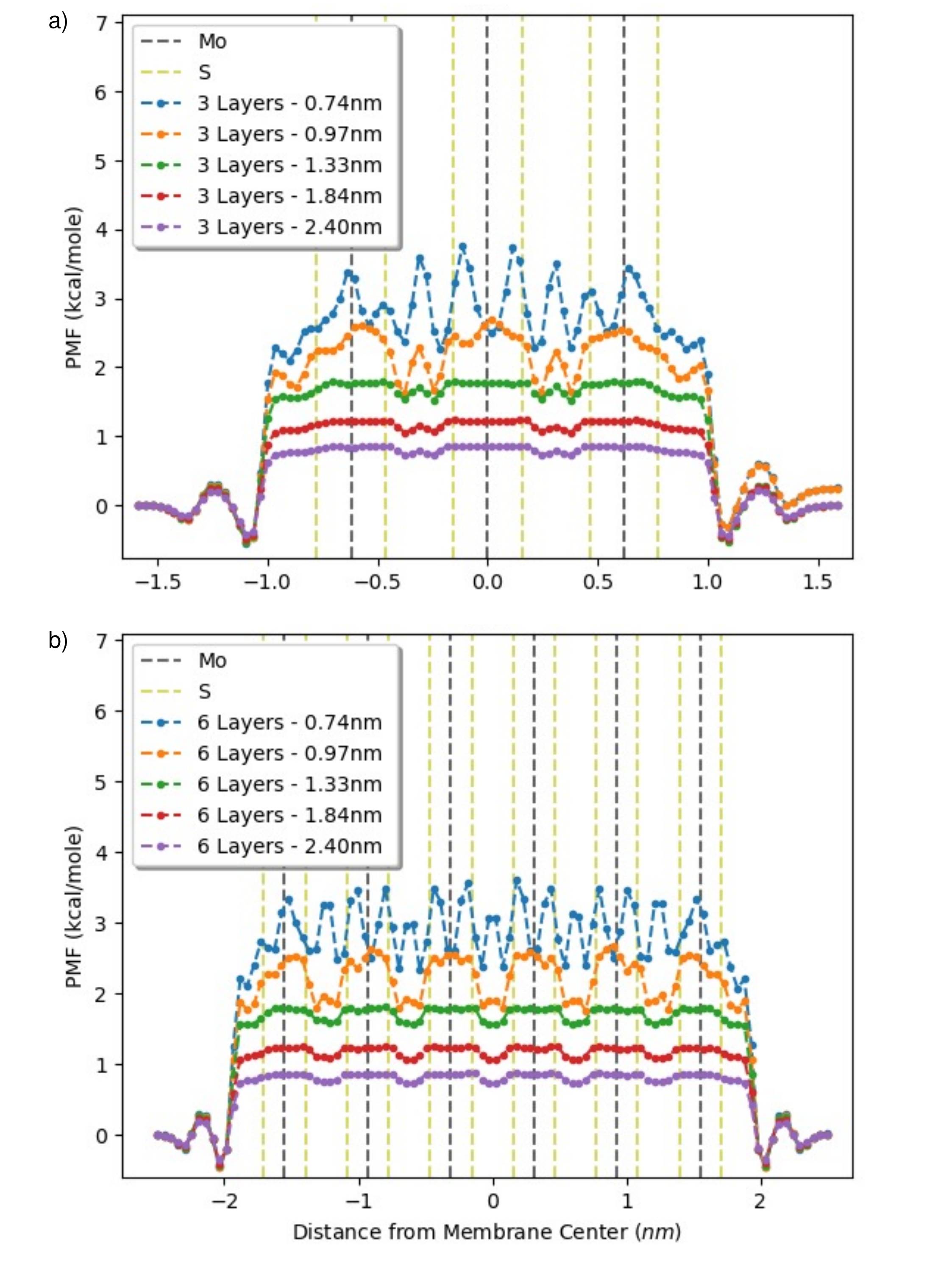}
\caption{The Potential of Mean Force (PMF) for (a) 3 layers and (b) 6 layers for each nanopore size.}
\label{FIG4}
\end{figure}

In order to understand the role played by the roughness in these systems we computed the PMFs as illustrated in Figure~\ref{FIG4}(a-b).
It shows that water molecules are trapped in the interlayer region, probably due to the change in the surface between water and the multi-layer.

\begin{figure}[!t]
\centering
\includegraphics[width=11.5cm]{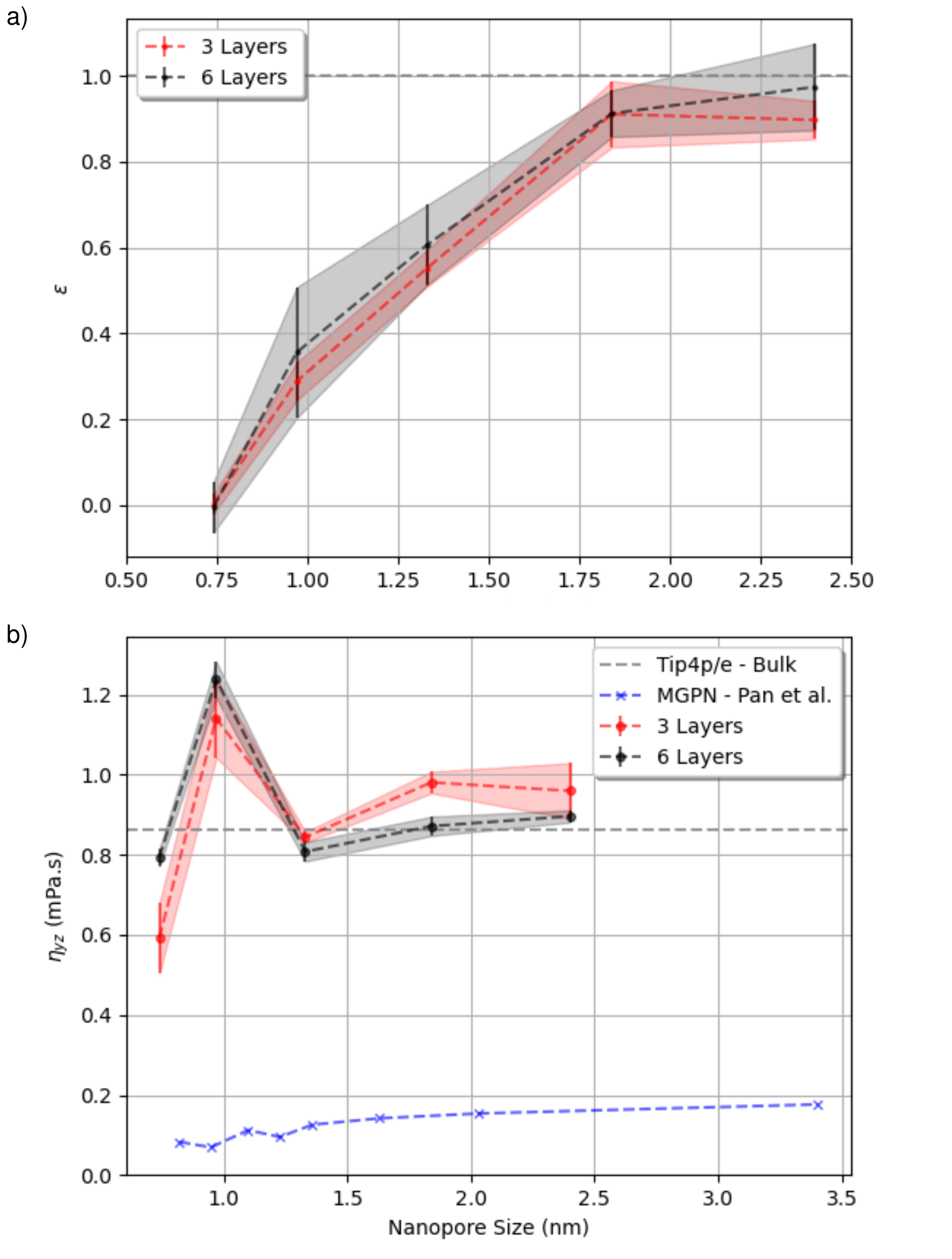}
\caption{(a) The flow enhancement factor $\epsilon$ {\it vs.} nanopore size for different number of layers.
(b) Shear viscosity {\it vs.} nanopore size.
Results for multi-layer graphene (MGPN) from reference~\cite{pan-jpcc2020} are depicted as blue marks.
The error bars are the deviation from the mean value (error bars smaller than the points are not shown).}
\label{FIG3-enha}
\end{figure}

In addition to the permeability, another important quantity to evaluate the performance of a membrane is the flow enhancement, $\epsilon$.
The concept was introduced originally for water confined in CNTs~\cite{hummer-nature2001, qin-nl2011}, where $\epsilon$ for nanotubes below 2 nm can exhibit values up to 1000.
It measures the mobility of confined water against the classical hydrodynamic prediction.
For our system, $\epsilon$ as a function of the nanopore size is shown in Figure~\ref{FIG3-enha}(a).
We can clearly observe that water flow is completely blocked in the smallest nanopore (see also Figures S1 and S6 of the Supporting Information) and monotonically increases with pore size, until the ratio $\epsilon = Q/Q_{HP}$ reaches 1 for the largest nanopore.
Pan et al. have observed that the enhancement factor in MGPN fluctuates around 1 for similar diameters~\cite{pan-jpcc2020}.
These values are at least one order of magnitude lower than that observed in CNTs~\cite{PhysRevLett.102.184502,doi:10.1021/nl200843g,doi:10.1063/1.4793396}.
The contrast between the behavior of $\epsilon$ for CNTs and our tube constructed from layered membranes can also be understood as follows.
While water flows almost stressless in contact with the carbon wall,
in our system the defects in the connections of the MLNMoS$_2$ membrane generates a slowing down in the dynamics.

In addition to the enhancement factor, the viscosity often offers elements to understand the dynamics of the fluid.
For instance, the viscosity of water confined in CNTs increases with the nanotube diameter as in the case of the multi-layer graphene, but it is one order of magnitude smaller than the MGPN~\cite{kohler-cpl2016}.
This can be attributed to the frictionless behavior of the water in nanotubes~\cite{kohler-ces2019},
whereas the assembly of the multi-layered graphene creates some friction.

In Figure~\ref{FIG3-enha}(b) we show the calculated shear viscosity for water inside both the 3 and 6-layer nanopores
and compare them to the results obtained for multi-layer graphene~\cite{pan-jpcc2020}.
The viscosity in the case of our system shows two distinctions when compared with the CNT and the MGPN systems:
it is much higher and it is not monotonic with the pore diameter.
In fact, it exhibits a large peak at the diameter of 0.97 nm.
The higher values are related to the unique nanopore architecture in MLNMoS2,
which acts on different fronts.
For instance, the chemistry composition and the resulting charge distribution are responsible for enhancing the electrostatic interaction with water.
Also, the pore geometry acts to increase friction at the interface, which contributes decisively to the increase in viscosity.
This is in agreement with the results for water flux and flow enhancement presented in Figure~\ref{FIG3-enha}(a),
showing a slowing down in the dynamics of water in MLNMoS$_2$ when compared to graphene~\cite{pan-jpcc2020}.
The steep increase in the viscosity for 0.97 nm can be understood in terms of the structural organization of the water molecules while passing through the nanochannel, which will be explored in the next section.

\subsection{The Anchor Effect}

\begin{figure}[!t]
\centering
\includegraphics[trim=0cm 0cm 0cm 0cm,clip, width=14cm]{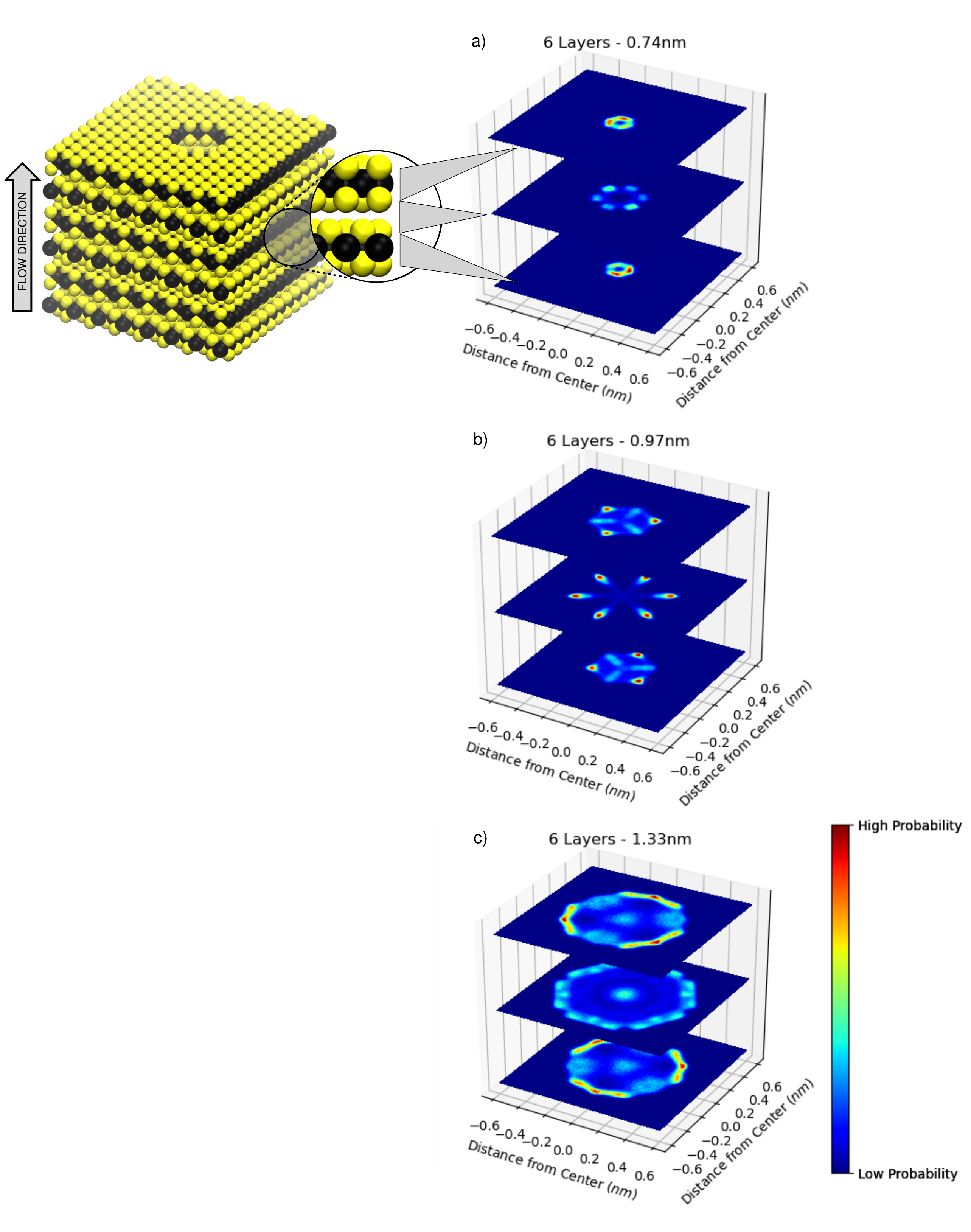}
\caption{3D visualization of the oxygen density map in between pores, the interlayer region (middle), and inside subsequent pores (top and bottom) for (a) 0.74, (b) 0.97 and (c) 1.33 nm diameter nanopores.}
\label{FIG5}
\end{figure}

The mechanism behind the low enhancement factor (Figure~\ref{FIG3-enha}(a)) and flow (Figure~\ref{FIG3}(a)) can be understood by examining the layered density color map of water oxygen atoms as illustrated in Figure~\ref{FIG5}.
The illustration shows the oxygen crossing the nanopore at three different regions:
two of them inside consecutive MoS$_2$ sheets (top and bottom) and one (the middle) in between those two sheets, the interlayer region.
We can see that water molecules assume very specific, expanded positions when traveling through the latter.
This structuring is evidenced by the six-folded star-like ring in Figure~\ref{FIG5}(b)-middle, where oxygen atoms become ``anchored'' in the interlayer region.
This interlayer obstacle present in the 0.74 nm, 0.97 nm, and 1.33 nm diameter nanopores result in the collective behavior of low flow,
low enhancement factor, and high viscosity when compared with nanotubes and MGPN.
Interestingly, the oxygen atoms inside the pore tend to be confined near to the Mo atoms due to the resulting dipole-dipole interaction between water molecules and the membrane~\cite{D1CP00613D}.

Another aspect which needs to be clarified is the large increase in the viscosity between 0.74 nm and 0.97 nanopore as shown in the Figure~\ref{FIG3-enha}(b).
Figure~\ref{FIG5} shows a clear difference  between the three interlayers.
For the 0.74 nm nanopore, water molecules are less organized in the interlayer region,
but they still exhibit an expanded arrangement due to the larger space available.
For the 0.97 nm pore, water in between MoS$_2$ layers gets organized.
The transition from the 0.74 nm to the 0.97 nm interlayer results in a steep change of the viscosity as illustrated in Figure~\ref{FIG4}(a).
As we increase the nanopore's diameter to 1.33 nm the difference between regions becomes less pronounced,
as is the case for the nanopores with larger diameters (Figure S4).

\begin{figure}[t!]
\centering
\includegraphics[trim=0cm 0cm 0cm 0cm,clip, width=11cm]{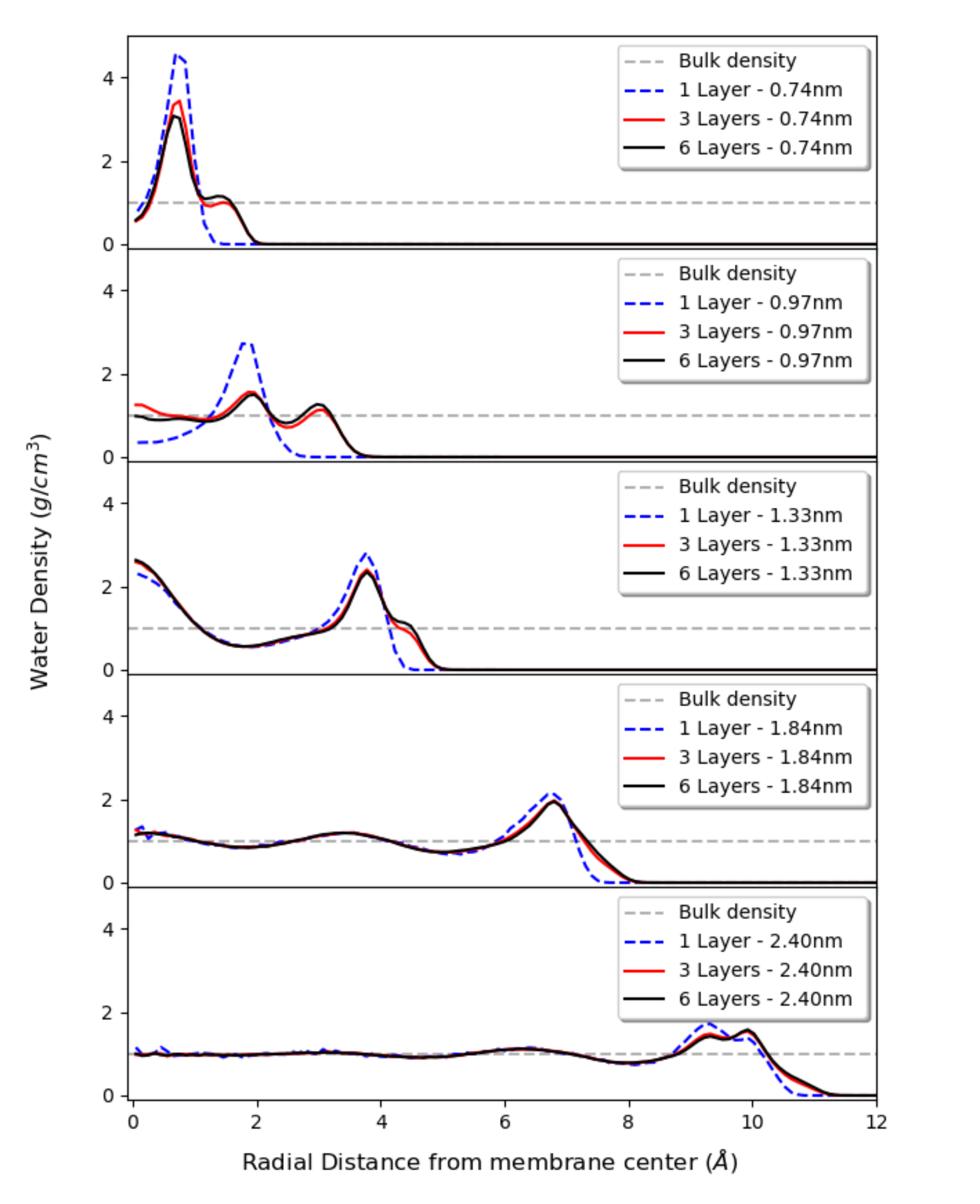}
\caption{Radial density profile (RDP) of water oxygen atoms inside 1, 3, and 6-layer nanopores.}
\label{FIG6}
\end{figure}

This change in the interlayer structure and its impact in water mobility can be further analyzed through the radial density profile of water.
In Figure~\ref{FIG6} we present the RDP of water molecules inside the nanopores.
The RDP is calculated by dividing the inner of the MoS$_2$ nanochannels in concentric cylindrical shells and averaging the number of oxygen atoms in each shell along the simulation.
We can observe a high structuring in the smaller nanopores ($\leq$1.33 nm) for both 1, 3, and 6-layer systems,
while bulk-like water can be seen at the center of the larger 2.4 nm channel.

To further confirm that water is trapped in the interlayer region, and therefore the anchor effect is hindering the water dynamics,
we plot in Figure S2 the difference between the channel water density (solid line), and the pore water density (dashed line) for the 6-layer case.
It is clearly shown that the water molecules occupy more expanded radial positions in the interlayer region than inside the MoS$_2$ nanopores.
The radial permanence time presented in Figure S2 (dotted line) confirms that water molecules tend to spend considerably longer times at these outer positions.
The phenomenon is more pronounced at the smaller 0.74 nm and 0.97 nm channels, as we can see in Figure S2(a),
which is a strong evidence that the interlayer region plays an important role trapping water molecules.

Alternative systems (with a different stacking symmetry, see Figure S7 of the Supporting Information)
were also implemented to observe how water structure and dynamics can be impacted by the nanochannel architecture.
Figure S8 shows that the water flow is even smaller in this case as compared to the 2H symmetry,
especially for the larger nanopores.
Additionally, The RDPs of Figure S9 shows that the same trend of expanded arrangements of water molecules are present in the alternative system.
This confirms the importance of the interlayer region to the water permeation in MLNMoS$_2$ membranes
and indicates that this phenomenon is present in different geometries and stacking symmetries.

As we can see, the increased organization of the water molecules as well as their expansion in the interlayer region (the anchor effect) are the main ingredients responsible for the lower flow and enhancement factor in Figure~\ref{FIG3-enha}(a).
In addition, the increase in the water viscosity when comparing the 0.74 nm with the 0.97 nm pore can be directly associated with the pronounced outer peaks appearing in Figure~\ref{FIG6} and S2.
This is the most dramatic example of the anchor effect, which prevents water from flowing freely through the channel.
Altogether, these results provide a solid explanation on the ubiquitous water flux in MoS$_2$ nanochannels and how it is dependent on several factors, such as the water local structuring due to the pore's chemistry and geometry as well as the interlayer spacing acting as a reservoir to slow down the flow.

\section{Conclusions}

In this paper we found a non-monotonic dependence of the water flux and water permeance on the thickness of MLNMoS$_2$ membranes.
There is a decrease in the water flux when more layers are considered,
but this decrease is neither proportional to the number of layers nor to the length of the nanochannel,
which violates the classical hydrodynamic prediction.

Even though the multi-layer design creates a nanochannel with some similarities to the CNT, we did not observe any enhancement factor in MLNMoS$_2$.
We offer an explanation for both phenomena in terms of the anchor effect that emerges due to the membrane's interlayer spacing,
which introduces a high degree of roughness to the nanochannel.

We also observed an anomalous higher water viscosity when compared to the graphene multi-layered system and a large increase in the viscosity when changing the nanopores from 0.74 nm to 0.97 nm.
This super viscous behavior is a result of an expanded structuration in the interlayer region.
These results provide useful information on the physics of the water flux in MoS$_2$ nanochannels and how it is dependent on the pore's chemistry and geometry as well as the interlayer spacing of the membrane.
They can also guide the advancement of current desalination technologies based on 2D membranes.

\section*{Supporting Information}

File (pdf) containing details on the water flux through the 2H and the alternative structure,
water radial density profiles and permanence time, color maps and system geometries.

\begin{acknowledgement}
This study was financed by Brazilian agencies, in part by the Coordenação de Aperfeiçoamento de Pessoal de Nível Superior (CAPES) - Finance Code 001 and in part by CNPq (Grant nr. 201097/2020-6) and INCT-FCx. We thank the CENAPAD-SP and CESUP-UFRGS for the computational time.
\end{acknowledgement}

\bibliography{REFERENCE}

\newpage

\section*{TOC Graphic}
\begin{figure}[h]
\centering
\includegraphics[width=8.5cm]{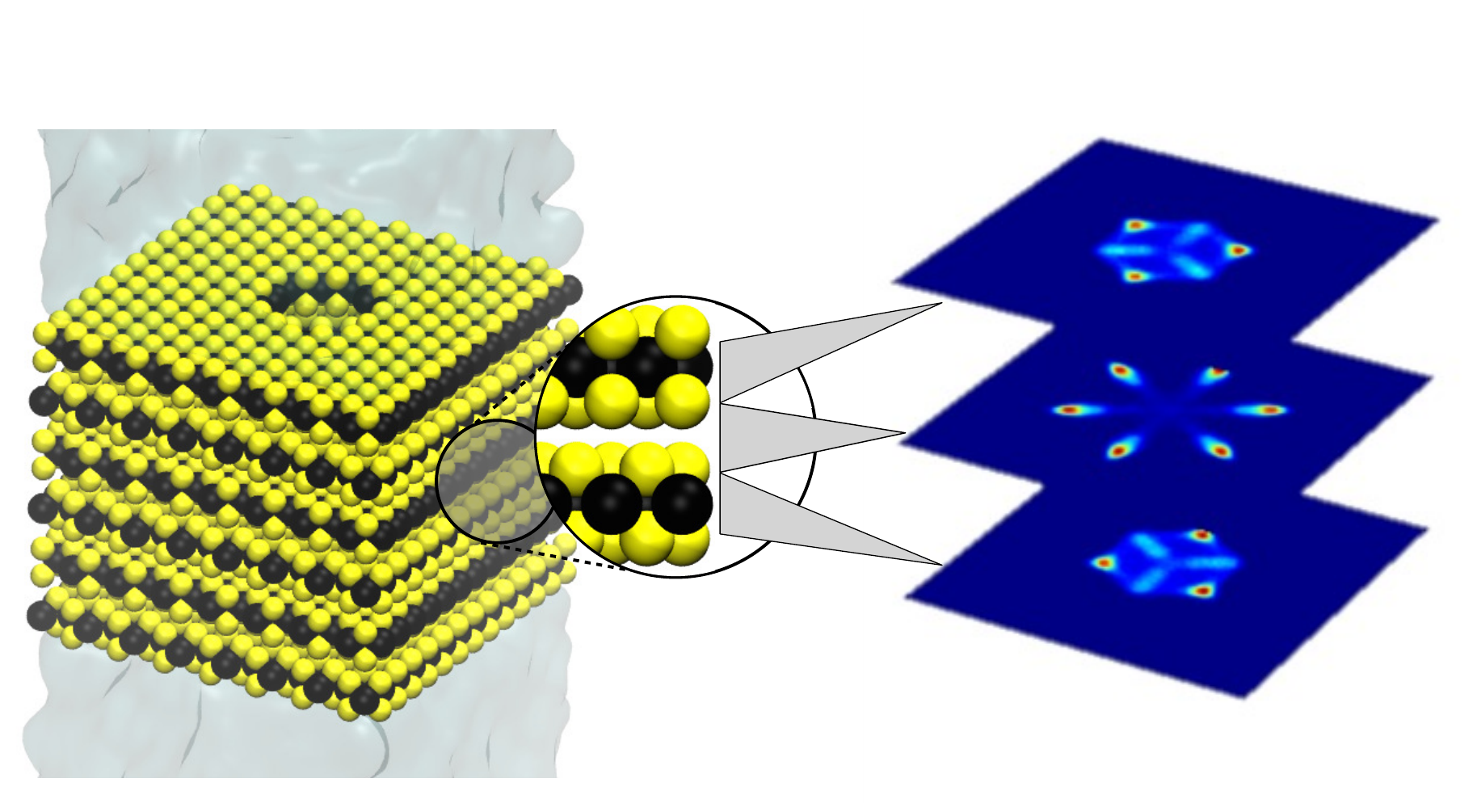}
\end{figure}
\end{document}